\documentclass[fleqn,twoside]{article}
\usepackage{espcrc2}

\newcommand{\psib}{\ensuremath{\overline{\psi}}}

\usepackage{latexsym}

\newcommand{\AmS}{{\protect\the\textfont2
  A\kern-.1667em\lower.5ex\hbox{M}\kern-.125emS}}

\title{Lattice Supersymmetry and Topological Field Theory}
\author{Simon Catterall\address{Department of Physics, Syracuse University,\\
Syracuse, NY 13244, USA}}

\begin{document}

\begin{abstract}We discuss the connection between supersymmetric field theories
and topological field theories and show how this connection may be used to
construct local lattice field theories which maintain an exact supersymmetry. 
It is shown how metric independence of the continuum topological field theory allows us to
derive the lattice theory by blocking out of the continuum in a deformed
geometry. This, in turn allows us to prove the cut-off independence of certain
supersymmetric
Ward identities.
\end{abstract}

\maketitle

\section{Motivation}

There have been many attempts to
formulate supersymmetric theories on lattices.
\cite{rev}.
But, unfortunately, generic lattice models break supersymmetry explicitly 
allowing relevant
SUSY violating operators to appear in the effective action.
The couplings of such operators must then be fine tuned to approach
a supersymmetric fixed point as the lattice spacing $a\to 0$.

Because of these problems one is motivated
to try to preserve an element of SUSY on the lattice in the hope
that this residual supersymmetry will 
protect the lattice theory from 
dangerous radiative corrections.
In this talk I will discuss one way to achieve this -- by exploiting a
well-known connection
between theories with extended supersymmetry and topological field theories
\cite{top}.
The prototype example of this is supersymmetric quantum mechanics viewed
as (0+1) dimensional lattice field theory. Another approach which also
attempts to find lattice actions exhibiting an exact supersymmetry can be
found in
\cite{kaplan}.

\section{SUSY QM}

Consider a model built from a discrete set of scalar fields $\phi_i$ 
with classical action $S_{\rm cl}(\phi)=0$. Such a model
is trivially invariant
under the shift (topological) symmetry
\begin{displaymath}
\phi_i\to \phi_i+\epsilon_i
\end{displaymath}
To quantize this theory we need to pick a gauge condition eg. $N_i(\phi)=0$.
The partition function of the gauge-fixed theory then takes the form
\begin{displaymath}
Z=\int D\phi e^{-\frac{1}{\alpha}N_i^2\left(\phi\right)} 
{\rm det}\left(\frac{\partial N_i}{\partial \phi_j}\right)
\end{displaymath}
If the determinant is represented using anticommuting ghosts 
$\psib_i$, $\psi_i$
and an auxiliary field $B_i$ is introduced we can rewrite this as
$$Z=\int D\phi D\psi D\psib DB e^{-S_q}$$
where
\begin{displaymath}
S_{\rm q}=-\frac{1}{2}\alpha B_i^2+N_iB_i+\psib_i\frac{\partial N_i}{\partial \phi_j}\psi_j
\end{displaymath}

$S_{\rm q}$ is then invariant under the BRST symmetry:
\begin{eqnarray*}
\delta \phi_i &=& \psi_i\xi \nonumber\\
\delta \psib_i &=& B_i\xi \nonumber \\
\delta \psi_i &=& 0 \nonumber\\
\delta B_i &=& 0
\end{eqnarray*}
We see that indeed that $\delta$ is a nilpotent operator on these fields.
Indeed, the action may be written
\begin{displaymath}
S_{\rm q}=\delta\left(\psib_i\left(N_i-\frac{\alpha}{2} B_i\right)\right)
\end{displaymath}
We will show in the next section that this structure is rather special and
allows us to write down expectation values {\it in the
continuum version of this model} whose values are independent
of the metric. These theories are thus termed topological quantum field theories TQFT.

Returning to our discrete model we can now make a specific choice of gauge
function and an associated physical interpretation of the field indices.
If the latter are
now chosen to label a series of lattice sites on a circle and the field $N_i$ chosen
as 
$$N_i=D^+_{ij}\phi_j + P^\prime_i(\phi)$$
where $D^+$ is the usual forward difference operator on the lattice
and $P^\prime$ is some arbitrary polynomial function of the field $\phi$, 
we can reinterpret the resulting theory as a lattice regulated version of supersymmetric
quantum mechanics in Euclidean space
\cite{us1}. Notice that this requires us also to
reinterpret the ghosts of the topological theory as the physical fermions
of the supersymmetric theory.  This is one aspect
of the way in which the TQFT and its SUSY parent differ.
The physical state conditions arising from the gauge fixing of the shift
symmetry ensure, in addition, that the true TQFT corresponds to the projection to the vacuum states of the parent SUSY theory.
In our construction we drop this projection and treat the discrete TQFT
action as a bona fide lattice action of the associated SUSY model possessing
an exact supersymmetry. Our numerical results indicate that this
is fine at least in low dimensions \cite{us2}. 

\section{Continuum TQFT}

The ingredients of a topological quantum field theory are a 
Riemannian manifold $g_{\mu\nu}$, a set of fields 
$\Phi$ and an action $S(\Phi)$ together with a special class of
so-called
topological observables $O\left(\Phi\right)$ which have the property 
\begin{displaymath}
\frac{\delta}{\delta g^{\mu\nu}}\left<O\right>=0
\end{displaymath}
This property can be guaranteed if the action is the variation of
some other function with respect to a nilpotent symmetry $S=\delta \Lambda$. 
The proof requires that both the measure and the
operator $O$ be invariant under the symmetry (and furthermore the latter
should not contain the metric explicitly)
It is important for our purposes to notice that any such theory
always contains a {\it trivial} set of topological observables -- namely
those functions which arise from the symmetry variation of some
other function
$O=\delta O^\prime$. In the language of supersymmetry these operators
generate the supersymmetric 
Ward identities of the parent SUSY theory

\section{Lattice Action as a Perfect Action}

This metric independence of the continuum theory may be exploited so
as to derive the lattice model directly from the continuum.
The lattice theory can be obtained by a process of blocking out of the
continuum in a deformed metric. 
We illustrate this in the 1D model we have
introduced.
In the continuum the bosonic action takes the form
\begin{displaymath}
S_B=\int dt\, e(t) \left[\frac{1}{e(t)}\frac{d\phi}{dt}+P^\prime(\phi)\right]^2
\end{displaymath} 
where $e(t)$ is the einbein
Change variables to {\it block} fields:
\begin{displaymath}
\phi^B(t)=\int dt^\prime\, e(t^\prime)B_\beta^-\left(t-t^\prime\right)\phi(t^\prime)
\end{displaymath}
where
\begin{displaymath}
B_\beta^-(t)=\frac{1}{2a}\left[L_\beta\left(t+\delta\right)-
L_\beta\left(t-a+\delta\right)\right]
\end{displaymath}
with
\begin{displaymath}
\lim_{\beta\to\infty}L_\beta(t)= \theta(t)
\end{displaymath}
Consider also an associated deformed metric
\begin{displaymath}
e_\beta(t)=\sum_{n=1}^N \frac{a}{A(\beta)} L_\beta^\prime(t-na)
\end{displaymath}
We can easily show that in $\lim_{\beta\to\infty}$
\begin{displaymath}
\phi^B(t)= \sum_{n=1}^N
\phi(na)\frac{1}{2A_L}\left[\theta(t-na)-\theta(t-(n+1)a)\right]
\end{displaymath}
Thus the field $\phi^B$ approaches a constant value within cells of a 1D
lattice -- changing only on moving from one cell to
another. 
Furthermore, its derivative takes the form
\begin{eqnarray*}
\lefteqn{\frac{1}{e_\beta(t)}\frac{d\phi^B}{dt}=\sum_{n=1}^N
\frac{1}{2aA_L}D^-_{nm}\phi_m \times} \\ &
&\left[ \theta(t-(n-1/2)a)-
        \theta(t-(n+1/2)a)\right]\\ \nonumber
\end{eqnarray*}
which is constant now within cells of a {\it dual} lattice.
Furthermore, the continuum bosonic action evaluated on such a 
continuum field resembles the lattice bosonic action
\begin{displaymath}
\lim_{\beta\to\infty}S_B(\phi^B)=\sum_{n=1}^N a
\left[\frac{1}{2A_La}D^-_{nm}\phi^B_m+P^\prime_n(\phi^B)\right]^2
\end{displaymath}
Similar arguments apply to the full action evaluated on both scalar and fermion
block fields. 
As $\beta\to\infty$ one can show
that $Z$ is dominated
by such block fields (the form of the scalar kinetic term ensures this) 
Thus the lattice theory, from the point of
view of topological observables, is merely the continuum
theory in the limit of a singular background and evaluated
using a new set of block variables. We would then expect
all the BRST symmetry of continuum theory to be
manifested at the quantum level on the lattice. Our simple model
affords an example of the latter feature.

SUSY QM actually has 2 topological symmetries in the continuum
which correspond to the flip $\psi\to \psib$ -- the 
original BRST symmetry  (now called $\delta_1$) and a second BRST
symmetry $\delta_2$. Now the lattice action is, by
construction, classically invariant under $\delta_1$ but {\it not}
under the second symmetry $\delta_2 S \ne 0$. The breaking term
would be a total derivative in the continuum but is
non-vanishing in general on the lattice. Nevertheless, as our
previous argument would have us believe, the Ward identities
following from $\delta_2$ are accurately satisfied on the lattice.
The tables below show numerical data for the bosonic and fermionic contributions
to Ward identities for both $\delta_1$ and $\delta_2$ 
for a simple model with $P^\prime=g\phi^3$ on a $N=4$ site lattice with
large lattice spacing.
 
\begin{table}[hbt]
\caption{$\delta_{(1)}\left(\psib_tx_0\right)$}
\label{table1}
\begin{tabular}{@{}lll}
\hline
t  & $<x(0)N^{(1)}(t)>$ & $<\psi(0)\overline{\psi}(t)>$  \\  
0 & 0.8895(11) & -0.8898(3) \\
1 & 0.6152(10) & -0.6155(3) \\
2 & 0.4294(11) & -0.4295(3) \\
3 & 0.3024(11) & -0.3028(3) \\
\hline      	
\end{tabular}
\end{table}

\begin{table}[hbt]
\caption{$\delta_{(2)}\left(\psi_tx_0\right)$}
\label{table2}
\begin{tabular}{@{}lll}
\hline
t  & $<x(0)N^{(2)}(t)>$ & $<\overline{\psi}(0)\psi(t)>$      \\
0 & -0.8895(11) & 0.8898(3) \\
1 & -0.3016(11) & 0.3028(3) \\
2 & -0.4294(11) & 0.4295(3) \\
3 & -0.6160(10) & 0.6155(3) \\
\hline      	
\end{tabular}
\end{table}

\section{Conclusions}

Certain models with extended SUSY 
can be related
to topological quantum field theories. The 
BRST charge(s) that appear in the latter theories
are formed from linear combinations of supercharges
(a procedure termed {\it twisting} in the literature) 
The topological symmetries may often be extended to the 
lattice and lead to models
which maintain some exact residual supersymmetry. Indeed, the algebra
$Q^2=0$ evades all the usual no-go theorems associated with lattice SUSY.
Furthermore, we have showed how the
lattice theories may be obtained by blocking out
of continuum in carefully chosen background geometry. This procedure
naturally generates a $r=1$ Wilson mass term for the fermions. 
Finally these ideas may be extended to 2D sigma models 
and Yang-Mills models in four dimensions which is work currently underway.


\begin{thebibliography}{9}
\bibitem{rev} A. Feo, Supersymmetry on the lattice, hep-lat/0210015.
\bibitem{top} S. Catterall, JHEP 0305 (2003) 038.
\bibitem{kaplan} A. Cohen, D. Kaplan, E. Katz and M. Unsal, hep-lat/0307012\\
                 A. Cohen, D. Kaplan, E. Katz and M. Unsal, JHEP 0308 (2003) 024.\\
		 D. Kaplan, E. Katz and M. Unsal, JHEP 0305 (2003) 037.
\bibitem{us1} S. Catterall and E. Gregory, Phys. Lett. B487 (2000) 349.
\bibitem{us2} S. Catterall and S. Karamov, Phys. Rev. D65 (2002) 094501.
\end{thebibliography}
\end{document}